\documentclass{bmcart}

\usepackage{amsthm,amsmath}
\usepackage[numbers,sort&compress]{natbib}
\usepackage[utf8]{inputenc} %unicode support
\usepackage{soul}
\usepackage{float}
\usepackage{placeins}
\usepackage{pdflscape}

\usepackage[para,online,flushleft]{threeparttable}
\usepackage{booktabs}% http://ctan.org/pkg/booktabs
\usepackage{authblk} % For nice title page
\usepackage[utf8]{inputenc}
\usepackage[titletoc,title]{appendix}
\usepackage{bm}
\usepackage{multicol} % For merging table columns.
\usepackage[numbers,sort&compress]{natbib}

\usepackage{xcolor}
\usepackage{soul}
\usepackage{longtable}
\usepackage{tabularray}
\usepackage{caption}
\usepackage{url}
\DeclareUnicodeCharacter{0302}{\^}
\usepackage[normalem]{ulem}

\usepackage{graphicx,float}

\usepackage{float}
\usepackage[para,online,flushleft]{threeparttable}
\usepackage{booktabs}% http://ctan.org/pkg/booktabs

\usepackage{subfloat}
\DeclareUnicodeCharacter{0302}{\^}

\graphicspath{{Images/}}

\startlocaldefs
\endlocaldefs

\begin{document}

%%% Start of article front matter
\begin{frontmatter}

\begin{fmbox}
\dochead{Simulation study}


\title{Evaluating the performance of Bayesian cumulative logistic models in randomised controlled trials: a simulation study}

%%%%%%%%%%%%%%%%%%%%%%%%%%%%%%%%%%%%%%%%%%%%%%
%%                                          %%
%% Enter the authors here                   %%
%%                                          %%
%% Specify information, if available,       %%
%% in the form:                             %%
%%   <key>={<id1>,<id2>}                    %%
%%   <key>=                                 %%
%% Comment or delete the keys which are     %%
%% not used. Repeat \author command as much %%
%% as required.                             %%
%%                                          %%
%%%%%%%%%%%%%%%%%%%%%%%%%%%%%%%%%%%%%%%%%%%%%%

\author[
  addressref={aff1,aff2},                   % id's of addresses, e.g. {aff1,aff2}
  corref={aff1,aff2},                       % id of corresponding address, if any
% noteref={n1},                        % id's of article notes, if any
  email={chris.selman@mcri.edu.au}   % email address
]{\inits{C.J}\fnm{Chris J.} \snm{Selman}}
\author[
  addressref={aff1,aff2},
  email={katherine.lee@mcri.edu.au}
]{\inits{K.J}\fnm{Katherine J.} \snm{Lee}}
\author[
  addressref={aff3,aff4}
]{\inits{S.Y.C}\fnm{Steven Y.C.} \snm{Tong}}
\author[
  addressref={aff5,aff6}
]{\inits{M.J.}\fnm{Mark} \snm{Jones}}
\author[
  addressref={aff1,aff7, aff8},
  email={robert.mahar@mcri.edu.au}
]{\inits{R.K}\fnm{Robert K.} \snm{Mahar}}
%%%%%%%%%%%%%%%%%%%%%%%%%%%%%%%%%%%%%%%%%%%%%%
%%                                          %%
%% Enter the authors' addresses here        %%
%%                                          %%
%% Repeat \address commands as much as      %%
%% required.                                %%
%%                                          %%
%%%%%%%%%%%%%%%%%%%%%%%%%%%%%%%%%%%%%%%%%%%%%%

\address[id=aff1]{%                           % unique id
  \orgdiv{Clinical Epidemiology and Biostatistics Unit},             % department, if any
  \orgname{Murdoch Children's Research Institute},          % university, etc
  \city{Parkville},  
   \state{VIC},
  \postcode{3052},
  \cny{Australia}                                    % country
}
\address[id=aff2]{%
  \orgdiv{Department of Paediatrics},
  \orgname{University of Melbourne},
  \city{Parkville},  
   \state{VIC},
  \postcode{3052},
  \cny{Australia}                                    % country
}

\address[id=aff3]{%
  \orgdiv{Department of Infectious Diseases},
  \orgname{University of Melbourne at the Peter Doherty Institute for Infection and Immunity},
  \city{Parkville},  
   \state{VIC},
  \postcode{3052},
  \cny{Australia}                                    % country
}
\address[id=aff4]{%
  \orgdiv{Victorian Infectious Diseases Service},
  \orgname{The Royal Melbourne Hospital, at the Peter Doherty Institute for Infection and Immunity},
  \city{Parkville},  
   \state{VIC},
  \postcode{3052},
  \cny{Australia}                                    % country
}

\address[id=aff5]{%
  \orgdiv{Sydney School of Public Health, Faculty of Medicine and Health},
  \orgname{University of Sydney},
  \city{Sydney},  
   \state{NSW},
  \postcode{2006},
  \cny{Australia}                                    % country
}
\address[id=aff6]{%
  \orgdiv{Wesfarmers Centre of Vaccines and Infectious Diseases},
  \orgname{Telethon Kids Institute},
  \city{Perth},  
   \state{WA},
  \postcode{6008},
  \cny{Australia}                                    % country
}
\address[id=aff7]{%
  \orgdiv{Centre for Epidemiology and Biostatistics, Melbourne School of Population and Global Health},
  \orgname{University of Melbourne},
  \city{Parkville},  
   \state{VIC},
  \postcode{3052},
  \cny{Australia}                                    % country
}
\address[id=aff8]{%
  \orgdiv{Methods and Implementation Support for Clinical and Health Research Hub},
  \orgname{University of Melbourne},
  \city{Parkville},  
   \state{VIC},
  \postcode{3052},
  \cny{Australia}                                    % country
}
%%%%%%%%%%%%%%%%%%%%%%%%%%%%%%%%%%%%%%%%%%%%%%
%%                                          %%
%% Enter short notes here                   %%
%%                                          %%
%% Short notes will be after addresses      %%
%% on first page.                           %%
%%                                          %%
%%%%%%%%%%%%%%%%%%%%%%%%%%%%%%%%%%%%%%%%%%%%%%

%%\note{Sample of title note}     % note to the article
%\note[id=n1]{Equal contributor} % note, connected to author

\end{fmbox}% comment this for two column layout

%%%%%%%%%%%%%%%%%%%%%%%%%%%%%%%%%%%%%%%%%%%%%%%
%%                                           %%
%% The Abstract begins here                  %%
%%                                           %%
%% Please refer to the Instructions for      %%
%% authors on https://www.biomedcentral.com/ %%
%% and include the section headings          %%
%% accordingly for your article type.        %%
%%                                           %%
%%%%%%%%%%%%%%%%%%%%%%%%%%%%%%%%%%%%%%%%%%%%%%%

\begin{abstractbox}

\begin{abstract} % abstract
\small	
\parttitle{Background} 
The proportional odds (PO) model is the most common analytic method for ordinal outcomes in randomised controlled trials. While parameter estimates obtained under departures from PO can be interpreted as an average odds ratio, they can obscure differing treatment effects across the distribution of the ordinal categories. Extensions to the PO model exist and this work evaluates their performance under deviations to the PO assumption.

\parttitle{Methods} 
We evaluated the bias, coverage and mean square error of four modeling approaches for ordinal outcomes via Monte Carlo simulation. Specifically, independent logistic regression models, the PO model, and constrained and unconstrained partial proportional odds (PPO) models were fit to simulated ordinal outcome data. The simulated data were designed to represent a hypothetical two-arm randomised trial under a range of scenarios. Additionally, we report on a case study; an Australasian COVID-19 Trial that adopted multiple secondary ordinal endpoints.

\parttitle{Results} 
The PO model performed best when the data are generated under PO, as expected, but can result in bias and poor coverage in the presence of non-PO, particularly with increasing effect size and number of categories. The odds ratios (ORs) estimated using the unconstrained PPO and separate logistic regression models in the presence of non-PO had negligible bias and good coverage across most scenarios. The unconstrained PPO model under-performed when there was sparse data within some categories.

\parttitle{Conclusions} 
While the PO model is effective when PO holds, the unconstrained and constrained PPO and logistic regression models provide unbiased and efficient estimates under non-PO conditions.

\end{abstract}

%%%%%%%%%%%%%%%%%%%%%%%%%%%%%%%%%%%%%%%%%%%%%%
%%                                          %%
%% The keywords begin here                  %%
%%                                          %%
%% Put each keyword in separate \kwd{}.     %%
%%                                          %%
%%%%%%%%%%%%%%%%%%%%%%%%%%%%%%%%%%%%%%%%%%%%%%


% MSC classifications codes, if any
%\begin{keyword}[class=AMS]
%\kwd[Primary ]{}
%\kwd{}
%\kwd[; secondary ]{}
%\end{keyword}

\end{abstractbox}
%
%\end{fmbox}% uncomment this for two column layout

\end{frontmatter}

%%%%%%%%%%%%%%%%%%%%%%%%%%%%%%%%%%%%%%%%%%%%%%%%
%%                                            %%
%% The Main Body begins here                  %%
%%                                            %%
%% Please refer to the instructions for       %%
%% authors on:                                %%
%% https://www.biomedcentral.com/getpublished %%
%% and include the section headings           %%
%% accordingly for your article type.         %%
%%                                            %%
%% See the Results and Discussion section     %%
%% for details on how to create sub-sections  %%
%%                                            %%
%% use \cite{...} to cite references          %%
%%  \cite{koon} and                           %%
%%  \cite{oreg,khar,zvai,xjon,schn,pond}      %%
%%                                            %%
%%%%%%%%%%%%%%%%%%%%%%%%%%%%%%%%%%%%%%%%%%%%%%%%

%%%%%%%%%%%%%%%%%%%%%%%%% start of article main body
% <put your article body there>

%%%%%%%%%%%%%%%%
%% Background %%
%%
\section*{Introduction}

An ordinal outcome comprises a set of monotonically ordered categories such that the distance between categories may not necessarily be meaningfully quantifiable. The multiple categories of the ordinal endpoint can capture clinical outcomes more meaningfully than binary endpoints \cite{dodd2020endpoints} as many distinct patient states can be incorporated into a single outcome. Ordinal outcomes have become common in randomised controlled trials (RCTs), with the widespread use of the World Health Organisation (WHO) Clinical Progression scale in COVID-19 studies increasing familiarity among researchers \cite{selman2024statistical, marshall2020minimal, mathioudakis2020outcomes}. 

There are many ways to approach the analysis of ordinal outcomes, the most common being the proportional odds (PO) model \cite{selman2024statistical}. The PO model is an extension of the logistic regression model to account for more than two categories and assumes that all covariates exert the same effect on the cumulative log-odds for each dichotomous cut-point of the ordinal outcome \cite{harrell2015regression}, known as the PO assumption. The target quantity of interest from this model is the proportional OR \cite{breheny2015proportional, abreu2008ordinal} which represents the relative odds of a better outcome in the treatment group compared with the control group. Desirable properties of the PO model are that it estimates a single target estimand, the model is palindromic invariant (that is, a reversal of the ordering of the categories changes the sign of the estimated regression coefficients), and it is invariant under collapsibility, that is, combining or removing categories does not affect the estimate of the treatment effect \cite{ananth1997regression}. 

A drawback of the PO model is that the PO assumption may not hold in practice \cite{lieberson1985improvement, kim2003assessing, fullerton2012proportional} in which case it will mask differences in the treatment effect across categories. A comparison between intervention groups using the PO model is most meaningful when the assumption of stochastic ordering holds, that is, the underlying cumulative distribution function of the outcome in the treatment arms do not intersect. This implies that the direction of the ORs based on each cut-point of the ordinal scale are all in the same direction. If stochastic ordering does not hold then the treatment is beneficial for some components of the ordinal outcome and harmful for others, in which case the PO assumption is clearly violated and this pattern would not be apparent with the PO model.

Another option is to fit a logistic regression model to a dichotomisation, or separate logistic regression models at each dichotomisation, of the ordinal scale. This disadvantage with this approach is that it reduces the  statistical power to detect clinically relevant treatment effects compared to the PO model if the PO assumption is true  \cite{armstrong1989ordinal}.

An alternative approach for comparing treatments across categories of an ordinal outcome is to use a partial proportional odds (PPO) model of which two variants have been proposed in the literature: the unconstrained and the constrained PPO model \cite{peterson1990partial}. The PPO model allows the PO assumption to be relaxed for some or all cut-points of the ordinal outcome and estimates an OR for each cut-point (if PO is not assumed for the covariate; otherwise a proportional OR is estimated). This approach may be advantageous over separate logistic regression models for each cut-point as it leverages the ordinal structure of the data by retaining the information contained in the category rankings, which logistic regression discards, and each cut-point OR is estimated in a single model rather than separate models.

Relative to the PO model, the PPO model is more flexible at the cost of estimating more parameters \cite{fullerton2012proportional, fullerton2021ordered}. The unconstrained PPO model allows the ORs at each cut-point to vary freely across any number of the cut-points for a subset $p-q$ of the $p$ exposure variables in the model (such that $q \leq p$) \cite{ananth1997regression}. The unconstrained PPO model allows a treatment to exert a different effect across the cumulative distribution of the ordinal outcome and an OR is produced for each cut-point. The drawback of this approach is that there is potential for negative probabilities to be estimated that arises when the model parameters are estimated in a way that does not align with the constraints imposed by the ordinal nature of the outcome \cite{fullerton2012proportional} (e.g.\ if the log-ORs lead to cumulative probabilities that violate the basic rules of probability, such as the cumulative probabilities decreasing with increasing cumulative logits).

Alternatively, the constrained PPO model allows the ORs for each cut-point to be modified by a common factor across some or all cut-points of the ordinal outcome \cite{abreu2008ordinal}. This common factor is a fixed scalar that depends on an assumed functional relationship between the cut-point log-ORs and cumulative logits, allowing the constrained PPO model to be more parsimonious and potentially more efficient than the unconstrained PPO model as less parameters are estimated. The constrained PPO model could be useful if, in advance, it is suspected that the treatment exerts a stronger or weaker effect only for the highest category in the ordinal outcome with the remaining log-ORs suspected to be equal \cite{ananth1997regression}. The disadvantage of the constrained PPO model is that misspecification of the functional relationship can affect the estimation of the target estimand (the cut-point OR).

The increased use of ordinal outcomes in RCTs during the COVID-19 pandemic underscores the need for a deeper understanding  of how various cumulative logistic models perform, particularly in scenarios where the assumption of PO may or may not be violated. This raises the question of whether more complex models, such as PPO models, offer any advantage over other cumulative logistic models in scenarios where a separate treatment effect for one or all cut-points is of interest. 

A recent scoping review of the use and analysis of ordinal outcomes in RCTs found that the PO model was the most commonly used analysis method. This was followed by dichotomising the ordinal scale and analysing the data using a standard method such as logistic regression \cite{selman2024statistical}. Alternative methods such as PPO models were rarely used. There has been some methodological research on the performance of PO models and the impact of estimand misspecification in the context of RCTs. For instance, a recent systematic review examined whether the PO model yields clinically meaningful treatment effects in COVID trials. The review found that the proportional OR could differ substantially from the OR estimated for clinically meaningful cut-points using a logistic regression model, especially when PO is violated \cite{uddin2023evaluating}. Simulation studies have demonstrated that analysing ordinal outcomes directly can substantially improve statistical power compared to dichotomising the outcome \cite{roozenbeek2011added, mchugh2010simulation, desantis2014regression}. However, there has been no comprehensive evaluation of the comparative performance of different cumulative logistic analytical approaches that can be used to estimate treatment effects in ordinal outcomes. 

The aim of this paper is to evaluate the relative performance of independent logistic regression models, the PO model, and the unconstrained and constrained PPO models through the implementation of a simulation study and a case study using the Australasian COVID-19 Trial (ASCOT) \cite{mcquilten2023anticoagulation}. We opted to use Bayesian models because they are flexible, offer many options for model regularisation where there are computational or inferential issues, and are becoming common in the analyses of ordinal outcomes (including the ASCOT case study).

We first formally describe the aforementioned analysis models that we use in our simulation study. We then describe our simulation study which uses a range of plausible data generating scenarios in the context of a two-arm RCT. We then describe the ASCOT study and apply the methods to the real-trial data. We conclude by summarising the key findings, strengths and weaknesses of this study and offer guidance and recommendations for practical use of the different modeling approaches.

\section*{Methods}
\subsection*{Notation}
In this section, we formally define the logistic, PO, and PPO models. For our purposes, we assume that the target parameter of inference is the log-OR(s) for treatment in the linear predictor for each model we considered. These parameters represents how much more likely a patient undergoing treatment is to be in category $k$ or higher compared to lower categories, relative to a control group. For the PO model, the log-OR are assumed to be constant across all cut-points, which we refer to as a proportional log-OR. We start by outlining our notation:

\begin{itemize}
    \item $j$ represents the total number of categories in the ordinal outcome;
    \item $\theta_k$ represents the log-OR for cut-point $k = 2, 3, ..., j$;
    \item $\beta$ represents the proportional log-OR;
    \item $ \gamma_{k}$ represents the increase in the increment associated with the $k{th}$ cumulative logit for $k = 2, 3, ..., j$ ;
    \item $\zeta$ represents the ratio of the cumulative log-odds of an individual being in the second category of higher for the treatment versus the control group (used in the unconstrained PPO model) ;
    \item $\Gamma_{k}$ represents the fixed scalar used in the constrained PPO model that depends on the assumed functional relationship between the cut-point log-ORs and each cumulative logit;
    \item $\gamma$ represents the fixed increase in the increment associated with each cumulative logit for a constrained PPO model, which is multiplied by $\Gamma_{k}$ ; 
    \item $\boldsymbol{\pi}_{i}$ is a vector of probabilities for the ordinal outcome in the control, $i = 0$, or treatment arm, $i = 1$. That is, $\pi_{i}=\left(\pi_{i 1}, \pi_{i 2}, \ldots, \pi_{i j}\right)$.
\item $\boldsymbol{\alpha} = (\alpha_{2},...,\alpha_{j})$ corresponds to a vector of `intercepts' in the model, which represent the cumulative log-odds of being in category $k$ or higher in the control group;
 \item $y_{a}$ represents the observed ordinal outcome for observation $a = 1, 2, ... , n_{obs}$.
 \item $x$ is the binary treatment indicator where 0 = control and 1 = treatment.
\end{itemize}

We next outline the different approaches for estimating the log-ORs across the categories of the ordinal outcome.

\subsection*{Analysis models}
\subsubsection*{Separate logistic regression models} 
A simple method that can be used to estimate the log-ORs for each cut-point is to fit separate logistic regression models to each binary split of the ordinal outcome where:

\begin{equation} \label{eq:logit}
\log \frac{P(Y = 1)}{P(Y = 0)}=\mathrm{logit}\big[P(Y = 1)] =\alpha +\theta_{k} \mathbf{x}
\end{equation}

where $k = 2,..., j$. The estimated treatment effects are the $\theta_k$'s from each of the separate logistic regression models, the ratio of the cumulative log-odds of being in category $k$ or higher (i.e., $Y = 1$) relative to the lower categories ($Y = 0$) for the treatment compared to the control group. Logistic regression is widely applied for the analysis of dichotomised ordinal outcomes and is an accessible method in practice, with implementation readily available in most statistical software. 

\subsubsection*{Proportional odds model}
The PO model estimates a proportional log-OR $ \beta$, for the treatment compared to the control, that is the treatment is assumed to exert a constant effect across each binary split of the ordinal scale. It assumes that each observation has an independent multinomial distribution, i.e..for $k = 2, 3, ..., j$:
\begin{equation} \label{eq:cumulative_logit}
\log \frac{P(Y \geq k)}{P(Y < k)}=\mathrm{logit}\big[P(Y \geq k)] =\alpha_{k}+ \beta \mathbf{x}
\end{equation}

\subsubsection*{Unconstrained partial proportional odds model}

The unconstrained PPO model allows the estimation of separate ORs for the treatment compared to the control across each cut-point in a single model. For $ k=2, \ldots, j$, the unconstrained PPO model can be written as:
\begin{equation} \label{eq:1}
    \mathrm{logit}\big[P(Y \geq k)] = \alpha_{k}+\mathbf{x} \boldsymbol{\zeta}+\boldsymbol{\mathrm{t}} \boldsymbol \gamma_{k}
\end{equation}

In Equation \ref{eq:1}, $\boldsymbol t$ is a $q \times 1$ vector of covariates in which PO is not assumed, and $\boldsymbol \gamma_{k}$ represents a $q \times 1$ vector of coefficients for the treatment indicator in $t$. Given the only covariate in this model is the treatment group, this implies that $q = 1$ and therefore the treatment covariate is PO is not assumed. In this model, $\mathrm{t} \gamma_{k}$ represents the increment that is associated with the $kth$ cumulative logit (for $k = 2,...,j$), such that $\gamma_{2} = 0$. This implies that $\boldsymbol \zeta$ represents the ratio of the cumulative log-odds of an individual being in the second category or higher for the treatment compared to the control group. The treatment effect for each cut-point is $\theta_k = \zeta + \gamma_{k}$.

\subsubsection*{Constrained partial proportional odds model} 

The constrained PPO model allows the treatment effect to vary across each binary split of the ordinal scale but constrained by way of a common factor. This common factor, denoted by $\Gamma_{k}$, is a fixed scalar that depends on the assumed functional relationship between the cut-point log-ORs and each cumulative logit. This scalar is multiplied by $\gamma$ in the calculation of each cumulative logit. For $ k=2, \ldots, j$, the constrained PPO model is:
\begin{equation}
 \mathrm{logit}\big[P(Y \geq k)] =\alpha_{k}+\mathbf{x}\boldsymbol{\zeta}+\mathbf{t} \boldsymbol{\gamma} \Gamma_{k},
\end{equation}

The fixed scalar for each cumulative logit is derived assuming the specified relationship between the cut-point log-ORs and cumulative logits (with  $\Gamma_{2} = 0$). For example, if there is suspected to be a linear relationship, then $\Gamma_{k} = k - 2$. The estimated treatment effect is represented by each $\theta_k = \zeta + \Gamma_{k} \gamma$, the cut-point log-OR for being in category $k$ or higher.

\section*{Simulation study}
We conducted a simulation study to evaluate and compare the performance of the above models under a range of distributional, proportionality, sample size, and effect size scenarios. In this section, we outline the data generating mechanisms, analysis methods, and performance measures used in the simulation study. 

\subsection*{Data generation}
\begin{figure}[h!] 
\centering
\caption{Schematic diagram of data generating mechanisms* }
\caption*{ \scriptsize *OR = odds ratio, PO = proportional odds }

\label{fig:fig1}
\advance\leftskip-3cm
\advance\rightskip-3cm
\includegraphics[width=1.1\textwidth]{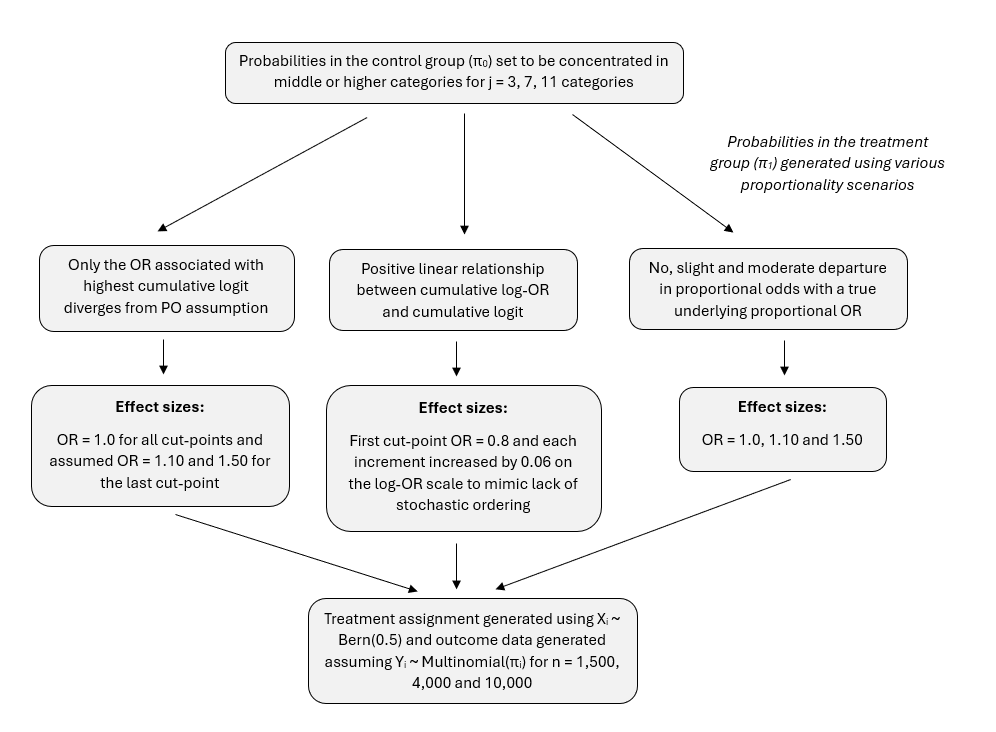}

\end{figure}

Data were generated under multiple scenarios which are summarised in Figure \ref{fig:fig1} for $n_{obs} = 1500, 4000$ and $10000$ observations for each scenario. These sample sizes were chosen to ensure the results are driven by the data rather than the prior distributions. We simulated different scenarios regarding proportionality which we describe further below, using $J = 3, 7$ and 11 categories. For each scenario, we first generated a treatment indicator $X_{i} \in \{0, 1\}$, where 0 = control and 1 = treatment, using $X_{i} \sim \mathrm{Bern}(0.5)$ representing simple randomisation with an equal allocation ratio. We obtained probabilities for the ordinal distribution in the control group, $\boldsymbol{\pi}_0$ using a $\theta \sim \mathrm{Beta}(\alpha, \beta)$ distribution, for which the probability density was discretised into $j$ distinct categories with the cut points for the $j$ bins evenly spaced across the distribution support. The values of $\alpha$ and $\beta$ were set to $\alpha, \beta = 1.8, 1.8$, for scenarios where the probability mass function is symmetric and concentrated in the middle, and $\alpha, \beta = 1.3, 0.9$ for scenarios where the probabilities in each category were negatively skewed (concentrated to the higher categories of the outcome). The treatment group distribution, $\boldsymbol{\pi}_1$, was then derived given the underlying (non-)proportionality scenario. The outcome data were then simulated by taking a random sample from a multinomial distribution for the control and treatment group separately using:

\begin{gather*}
    Y_0 \sim \mathrm{Multinomial}(\boldsymbol \pi_0) \\ 
    Y_1 \sim \mathrm{Multinomial}(\boldsymbol \pi_1)
\end{gather*}

The probabilities of each outcome distribution were therefore drawn from three different proportionality scenarios:

\begin{enumerate}
    \item \textbf{Random variability around a proportional log-OR}: We generated data under null (log(1)), weak (log(1.10)) and moderate (log(1.50)) effect sizes for the proportional log-OR, with varying degrees of non-proportionality around the proportional OR across the categories. This was achieved by generating outcome data from the model $\mathrm{logit}(P(Y\geq k) = \alpha_{k} + \theta_{k}$, where $\theta_{k}$ was randomly sampled from a truncated normal distribution with a mean of log(1), log(1.10) and log(1.50) and bounded above by $\theta_{k+1} - \alpha_k$ to avoid the estimation of negative probabilities. The standard deviation ($\sigma$) of the truncated normal distribution reflected the extent of the departure from PO using (1) no departure ($\sigma$ = 0), (2) slight departure ($\sigma$ = 0.05), and (3) moderate departure ($\sigma$ = 0.10), the latter two scenarios representing non-PO scenarios. The probabilities in each category in the treatment group, $\boldsymbol \pi_t$, were then computed using the inverse cumulative logit function. 

    \item \textbf{Linear relationship between cumulative logit and each cut-point log-OR}: In this scenario we generated data assuming a linear trend between the cut-point log-ORs and each cumulative logit such that $\Gamma_{k} = k - 2$ for $k = 2,...,j$, and $\Gamma_2 = 0$. Under this scenario, we specified a fixed cut-point log-OR between the treatment and control group that linearly increased for each cumulative logit, using $\gamma = 0.06$ and $\zeta =$ log$(0.8)$ (for the constrained PPO model) which reflects a scenario where the treatment exerts both a beneficial and harmful effect. The treatment cumulative log-odds were then calculated using $\mathrm{logit}(P(Y\geq k) = \alpha_{k} +$ log(0.8)$\times (k-2)$, which were back-transformed to a probability.

    \item \textbf{Allowing the increment associated with the highest cumulative logit to diverge from PO}: In this scenario we specified divergence from PO for the highest cumulative logit, forcing a proportional log-OR across the first $j - 2$ cumulative logits that represented no treatment effect. We specified a fixed cut-point $\mathrm{log(OR)}_{j} = \mathrm{log}(1.10)$ and $\mathrm{log}(1.50)$ between the treatment and control group for the highest cumulative logit, and the remaining cumulative logits had a fixed log-OR such that $\mathrm{log}(\theta_k) = 0$. Therefore the treatment cumulative log-odds were calculated using $\mathrm{logit}(P(Y\geq k) = \alpha_{k} + \theta_k$, where $\theta_k = log(1)$ for the first $j-2$ cumulative logits and $\mathrm{log(OR)}_{j} = \mathrm{log}(1.10)$ or $log-OR_{j} = \mathrm{log}(1.50)$ for the $j-1$ (highest) cumulative logit. These were then back-transformed to a probability.
\end{enumerate}

In total, 162 scenarios were considered in this simulation study. 

\subsection*{Analysis methods} 
Each dataset was analysed using separate logistic regression models, a PO model, and the unconstrained and two constrained PPO models (one assuming a linear functional relationship between each cumulative logit and log-OR, and the other assuming only the highest log-OR diverges from the PO assumption) all fitted using a Bayesian approach. We assigned minimally-informative Dirichlet implicit prior for the intercepts that is related to the probability of falling in a category for an individual $i$ at treatment means, i.e. $\boldsymbol \pi_i \sim \mathrm{Dirichlet}(\boldsymbol{1})$. This implies a uniform prior belief that is equivalent to one observation in each of the $J$ ordinal categories when the predictors are at their sample means, indicating very weak prior knowledge that no category has probability zero. Weakly informative normal priors centred at zero were used for the proportional and non-PO parameters in each cumulative logistic model to minimise prior influence, ensuring results were driven primarily by the data and underlying model assumptions. All methods of analysis were fitted using Markov Chain Monte Carlo approaches, using Hamiltonian Monte Carlo (HMC) sampling to obtain the posterior distribution. The HMC method was implemented in the software package R and Stan \cite{carpenter2017stan}, using the \textit{rmsb} package \cite{rmsb} and \textit{posterior} package \cite{posterior} for inference. The \textit{rmsb} package uses QR re-parametrisation to help the Markov chain move efficiently without compromising numerical accuracy. 

An iterative process was used to determine the number of simulations, $n_{sim}$ for each scenario, to ensure that the required maximum tolerable upper bound of the Monte Carlo standard errors (MCSEs) for each performance measure was less than 0.05 \cite{kelter2023bayesian}. We calculated the jackknife-after-bootstrap MCSE for each parameter using the method described in Koehler et al \cite{koehler2009assessment}. All scenarios achieved the maximum tolerable upper bound when $n_{sim} = 1000$.

A warmup of $3750$ iterations and a chain length of $7500$ samples in each chain were used for a total of 4 chains. No thinning was applied and the total post-warmup draws was $15000$ samples from the posterior distribution. Convergence diagnostics, including the effective sample size (ESS), potential-scale reduction factor, $\hat{R}$, and the number of divergent transitions (which indicate issues with exploring the parameter space properly), were recorded to assess the reliability of the posterior distributions. The number of divergent transitions were minimised or eliminated in the analysis by increasing the average target acceptance probability rate and the maximum tree depth. 

\subsection*{Performance metrics}

We compared the estimation performance of the various models using bias (using the median of the posterior distribution of the log-OR), relative bias (on the OR scale), 95\% credible interval coverage, and mean square error (MSE) of the log-OR for each cut-point.

\subsection*{Simulation study results}
\subsection*{Proportionality scenario 1: random variability around a proportional log-OR}

There was negligible bias in the estimated log-ORs across the majority of the cut-points for all analysis methods, which was an expected result given that the data were generated under an assumption of PO (Figure 2A). There was a small bias in the OR for the first cut-point from the unconstrained PPO and separate logistic regression models (relative bias $< 5\%$) when OR = 1.50, which was more pronounced when there was a skewed distribution across the ordinal categories in the control group, and did not appear to change with increasing variability about PO across the cut-points. Notably, the linear constrained PPO model exhibited greater bias for increasing variability in PO as the number of categories increased (Figure 7; Supplementary Material 1). 

The coverage probability was close to the nominal level (95\%) when data were generated with a true underlying proportional OR (Figure 2B) for all models. With increasing variability about PO (Figures 13-18; Supplementary Material 1), coverage slightly decreased compared to the nominal level for all analysis methods but varied across the cut-points. Coverage appeared to be higher in the extreme cut-points for the logistic and unconstrained PPO models, although the opposite relationship was observed for the linear constrained PPO model. Coverage across all cut-points decreased with increasing sample size for all models.  

For all scenarios where probabilities were generated using PO or a small variation about PO, the MSE for each cut-point was lowest for the PO model (Supplementary Material 1, Figure 28). Although the MSE was also close to zero for the logistic and unconstrained PPO models for the central cut-points, it was higher for the first cumulative logit when there were skewed outcome probabilities across the ordinal scale and for both extreme cut-points when there were symmetric outcome probabilities likely due to the lower event rates at the extremes. The MSE improved with increasing sample size for all models.

\subsection*{Proportionality scenario 2: linear relationship between cumulative logit and cut-point log-OR}

There was negligible bias in the ORs estimated from the linear constrained PPO, unconstrained PPO and separate logistic regression models across all cut-points and scenarios ($<2\%$ relative bias, see Figure 3A). There was, however, bias in the ORs estimated from the PO and the constrained PPO model that allowed the last OR to diverge from PO; this bias increased with a higher number of categories and when the outcome probabilities in the control group were skewed.

The linear constrained PPO model, separate logistic regression and unconstrained PPO models had coverage probabilities that were close to the nominal level (Figure 3B). However, there was under-coverage for the PO model and constrained PPO model allowing the last OR to diverge from PO for the scenarios with seven and 11 categories, particularly for the extreme cut-points and when the outcome probabilities in the control group were skewed. 

The linear constrained PPO yielded the lowest MSE across all scenarios (Supplementary Material 1, Figure 37), closely followed by separate logistic models and unconstrained PPO model, although the MSEs were larger in the extreme ends of the ordinal scale for these latter two methods.

\begin{landscape}
\begin{subfigures}

\begin{figure}[h!]
\centering
\caption{Relative bias in the cut-point odds ratios for each category of the ordinal outcome when data were generated under proportional odds }
\small\textsuperscript{OR = odds ratio, PO = proportional odds, PPO = partial proportional odds, LR = logistic regression, CPPO = constrained partial proportional odds}

\advance\leftskip-3cm
\advance\rightskip-3cm
\includegraphics[width=2.1\textwidth]{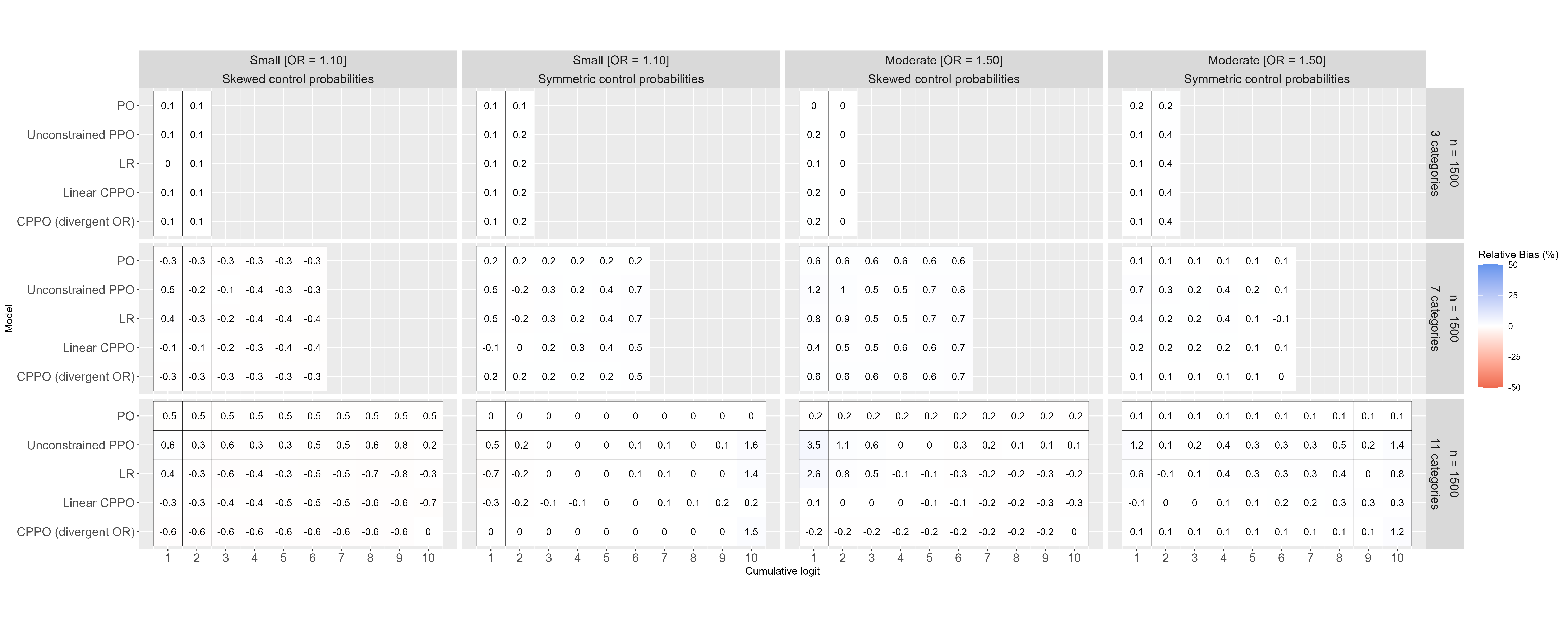}
\end{figure}

\begin{figure}[h!]
\centering
\caption{Coverage in the cut-point log-odds ratios for each category of the ordinal outcome when data were generated under proportional odds }
\small\textsuperscript{OR = odds ratio, PO = proportional odds, PPO = partial proportional odds, LR = logistic regression, CPPO = constrained partial proportional odds}

\advance\leftskip-3cm
\advance\rightskip-3cm
\includegraphics[width=2.1\textwidth]{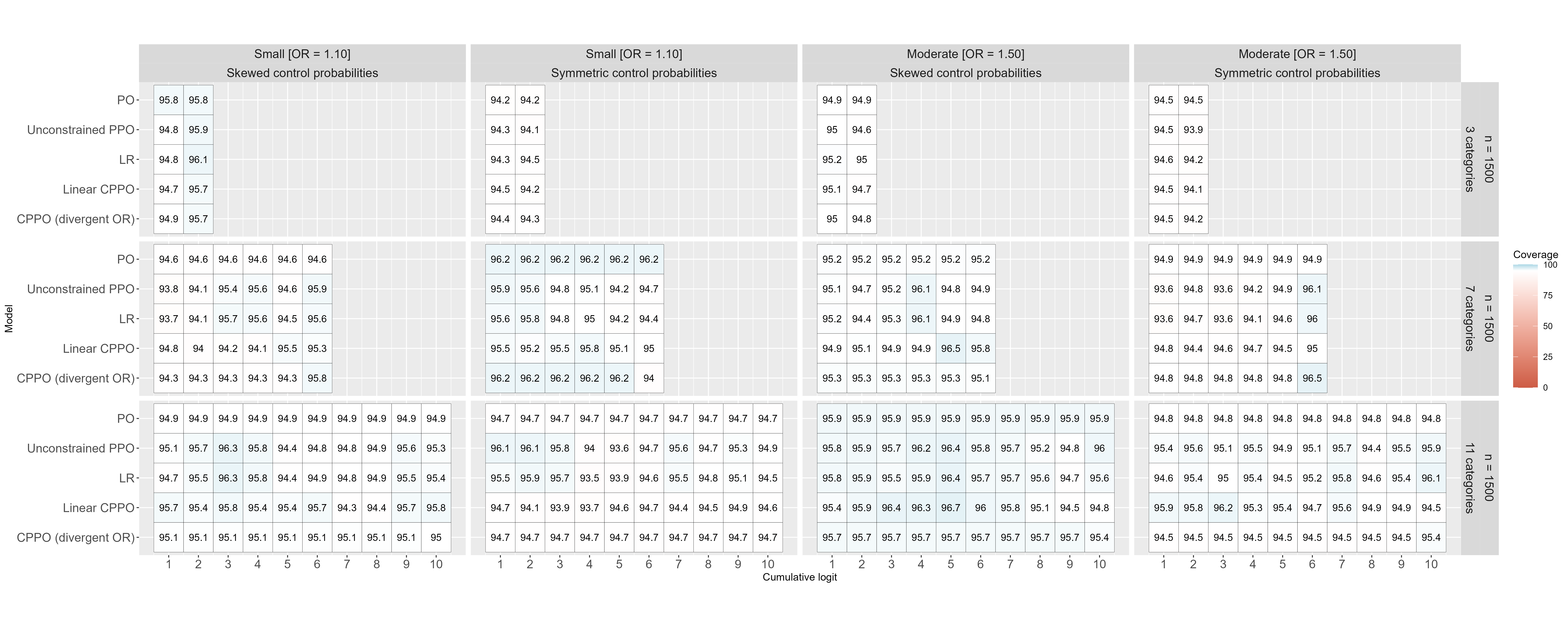}

\end{figure}

\end{subfigures}
\end{landscape}

\subsection*{Proportionality scenario 3: allowing the increment associated with the highest cumulative logit to diverge from PO}

Unsurprisingly, the constrained PPO model that allowed the increment associated with the highest cumulative logit to diverge from PO resulted in negligible bias (relative bias $< 1\%$) in proportionality scenario 3 as this was consistent with the data generating mechanism. The logistic and unconstrained PPO models also had negligible bias, although there was a small bias in the OR for the extreme cut-points (Figure 4A). The PO model resulted in high relative bias across the cut-points when there was a larger effect size. The bias was minimal across the first $j-2$ cumulative logits in scenarios where the control outcome probabilities were symmetric. However, there was bias in the cut-point OR for these categories in scenarios where the outcome probabilities in the control group were skewed. This bias was higher in the $j-1$ (highest) cumulative logit and increased when there was a higher number of categories in the ordinal outcome. 

Coverage remained roughly within $ \pm 2\%$ of the nominal level for most methods under most scenarios, particularly when the effect size was small (Figure 4B). However, where there were a large number of categories and a larger effect size, coverage decreased for the PO model for the highest cumulative logit, dropping close to 0\%. In contrast, the constrained PPO model allowing the highest cumulative logit to diverge from PO and the unconstrained PPO and separate logistic models resulted in coverage close to the nominal level (range 93.8\% to 96\%). 

\subsection*{Convergence Diagnostics}

The posterior distributions for all target parameters across all scenarios appeared to converge to their stationary distribution, with $\hat{R} < 1.01$ for all parameters. The bulk and tail effective sample sizes were above the recommendation of 100 per chain for all analytic approaches for all scenarios \cite{goodrich2020rstanarm,vehtari2021rank}, indicating all posterior distributions were informed by the data and there was low autocorrelation in the samples. The MCSE of all performance measures for all approaches/scenarios were less than $0.05$ and are reported in Supplementary Material 2. Finally, there were nine scenarios where there were divergent transitions. A sensitivity analysis was conducted that removed these simulated datasets from the analysis of the performance measures that yielded effectively identical results (see Supplementary Material 3).

\begin{landscape}
\begin{subfigures}
\begin{figure}[h!]
\centering
\caption{Relative bias in the cut-point odds ratios for each category of the ordinal outcome when data generated assuming a linear relationship between cumulative logit and cut-point log-OR}
\small\textsuperscript{OR = odds ratio, PO = proportional odds, PPO = partial proportional odds, LR = logistic regression, CPPO = constrained partial proportional odds}

\advance\leftskip-3cm
\advance\rightskip-3cm
\includegraphics[width=1.6\textwidth]{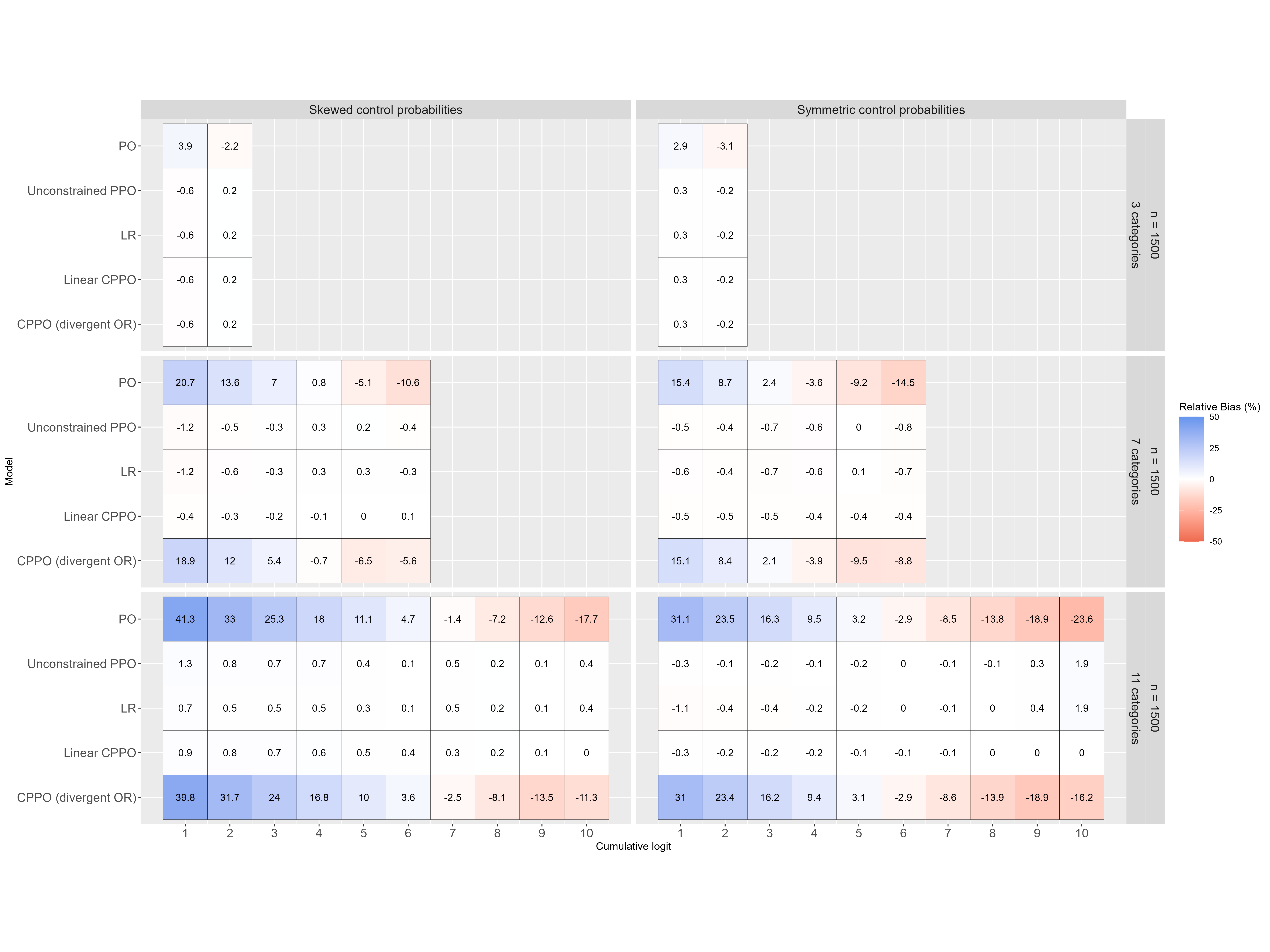}

\end{figure}

\begin{figure}[h!]
\centering
\caption{Coverage in the cut-point log-odds ratios for each category of the ordinal outcome when data generated assuming a linear relationship between cumulative logit and cut-point log-OR}
\small\textsuperscript{OR = odds ratio, PO = proportional odds, PPO = partial proportional odds, LR = logistic regression, CPPO = constrained partial proportional odds}

\advance\leftskip-3cm
\advance\rightskip-3cm
\includegraphics[width=1.6\textwidth]{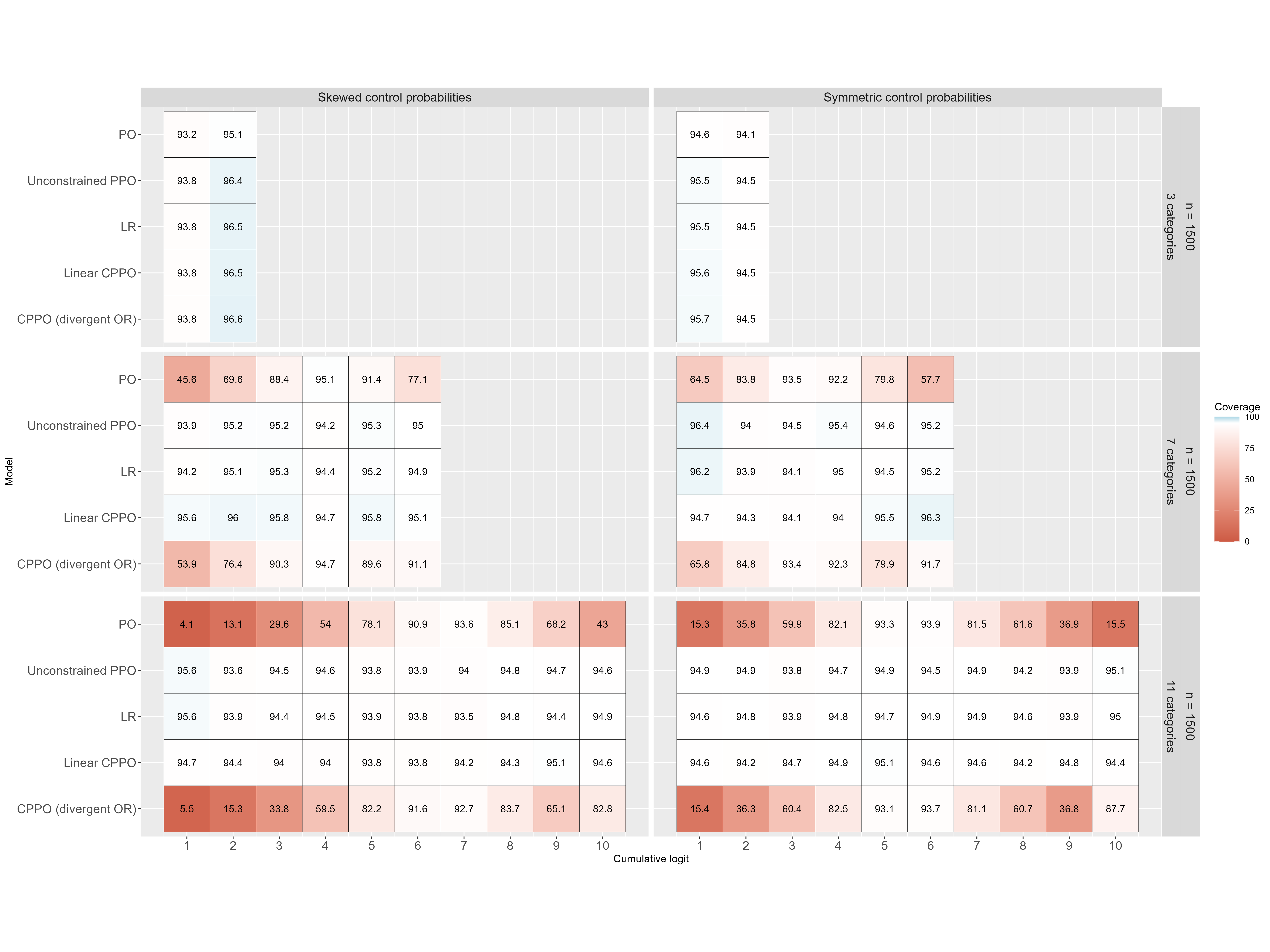}

\end{figure}

\end{subfigures}

\begin{subfigures}
\begin{figure}[h!]
\centering
\caption{Relative bias in the cut-point odds ratios for each category of the ordinal outcome when data generated assuming increment associated with highest cumulative logit diverges from proportional odds}
\small\textsuperscript{OR = odds ratio, PO = proportional odds, PPO = partial proportional odds, LR = logistic regression, CPPO = constrained partial proportional odds}

\advance\leftskip-3cm
\advance\rightskip-3cm
\includegraphics[width=2.1\textwidth]{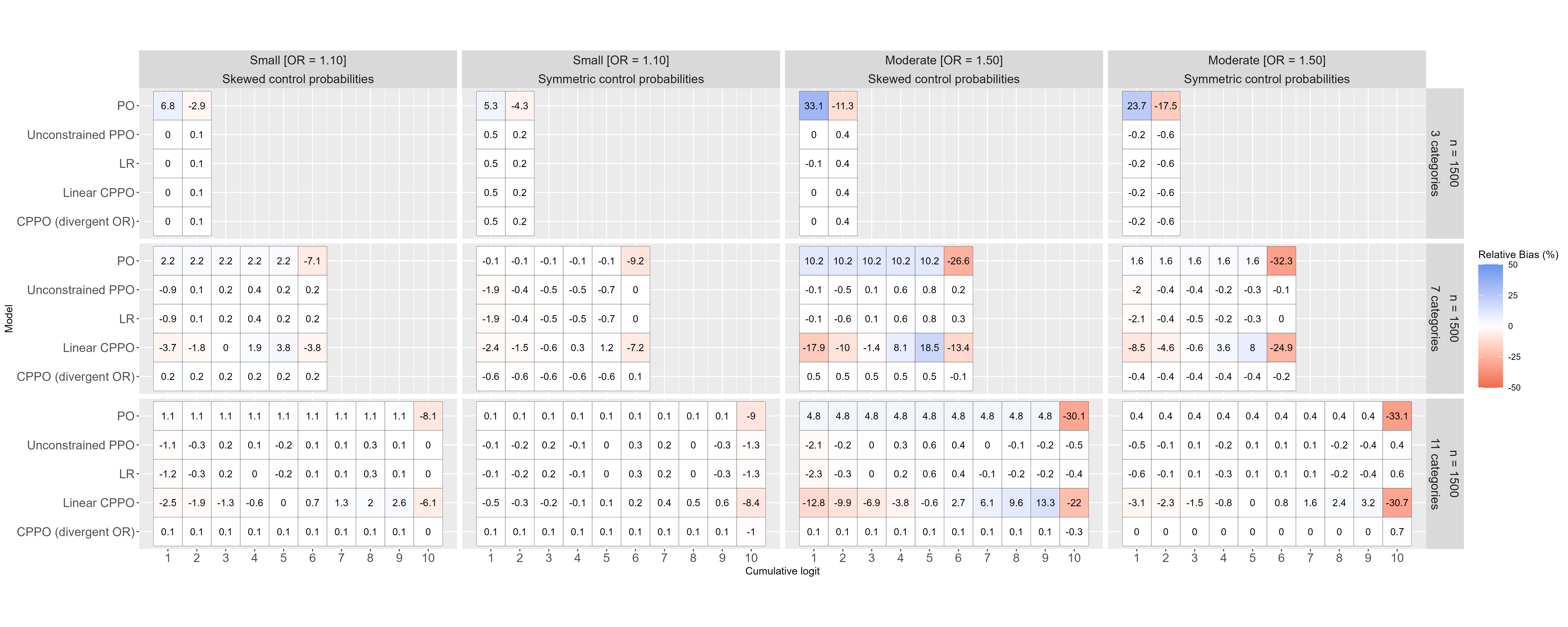}

\end{figure}

\begin{figure}[h!]
\centering
\caption{Coverage in the cut-point log-odds ratios for each category of the ordinal outcome when data generated assuming increment associated with highest cumulative logit diverges from proportional odds}
\small\textsuperscript{OR = odds ratio, PO = proportional odds, PPO = partial proportional odds, LR = logistic regression, CPPO = constrained partial proportional odds}

\advance\leftskip-3cm
\advance\rightskip-3cm
\includegraphics[width=2.1\textwidth]{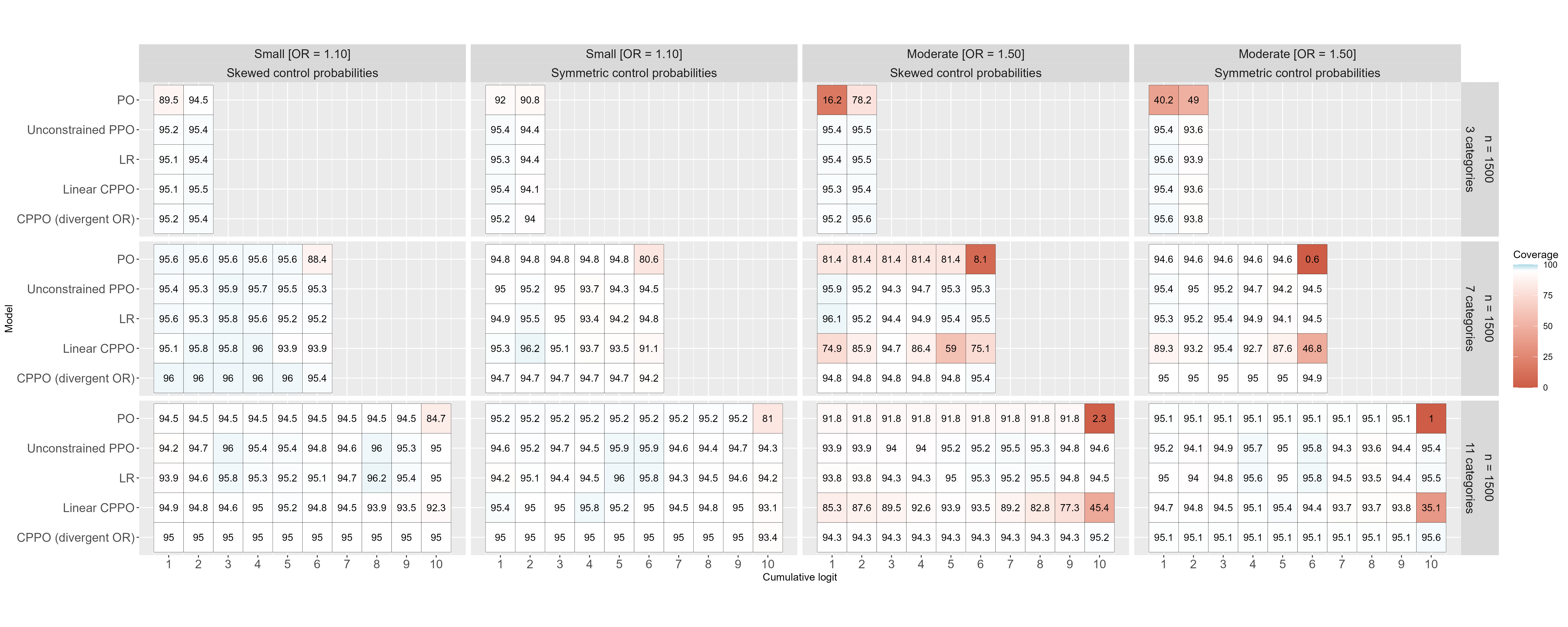}

\end{figure}
\end{subfigures}
\end{landscape}

\subsection*{Case study: The Australasian COVID-19 Trial (ASCOT)}
ASCOT is a platform trial that investigated multiple treatments for hospitalised, non-critically ill patients with COVID-19. ASCOT compared interventions in three treatment domains (anticoagulation, antiviral, and therapeutic antibody domains) each of which evaluated a set of treatments sharing a common modality. We restrict our focus to the anticoagulation domain which investigated a low-dose thromboprophylaxis with low-molecular-weight heparin (LMWH), intermediate-dose with LMWH, and low-dose with LMWH plus once-daily aspirin (the control). The primary outcome was death or the requirement for new organ support by day 28, a binary outcome. The low dose aspirin arm was removed part way through the trial based on external evidence, and a new therapeutic dose anticoagulation arm was added. The trial was stopped early because of funding constraints, slowing recruitment, and the data safety and monitoring committee recommending to stop the therapeutic anticoagulation treatment for futility. 

ASCOT had four ordinal secondary endpoints:

\begin{enumerate}
    \item WHO 8-point ordinal outcome scale at day 28 post randomisation: (1) not hospitalised and no limitations on activities, (2) not hospitalised, limitations on activities, (3) hospitalised, not requiring supplemental oxygen and no longer requiring ongoing medical care, (4) hospitalised, not requiring supplemental oxygen but requiring ongoing medical care, (5) hospitalised, requiring supplemental oxygen, (6) hospitalised, on non-invasive ventilation or high flow oxygen devices, (7) hospitalised, on invasive mechanical ventilation or ECMO, and (8) death.
    \item Modified Medical Research Council (mMRC) breathlessness scale (only asked among those diagnosed with COVID-19): 5-point ordinal scale  (1) `I only get breathless with strenuous exercise' (Grade 1), (2) `I get short of breath when hurrying on level ground or walking up a slight hill', (3) `On level ground, I walk slower than people of the same age because of breathlessness, or I have to stop for breath when walking at my own pace on the level', (4) `I stop for breath after walking about 100 yards or after a few minutes on level ground' to (5) `I am too breathless to leave the house or I am breathless when dressing or undressing' (Grade 5).
    \item Days alive and free of hospital by 28 days post randomisation: 29-point ordinal outcome calculated as 28 minus the number of days spent in hospital, where patients dying within 28 days were assigned zero free days. 
    \item Days alive and free of ventilation by 28 days post randomisation: 29-point ordinal outcome calculated as 28 minus the number of days on invasive or non-invasive ventilation, where patients dying within 28 days were assigned zero free days. 
\end{enumerate}

We focus on the first two endpoints in the manuscript and consider the others in the supplementary material.

\section*{Analysis of the case study}

The same analytic methods were applied to the 4 ordinal secondary outcomes in ASCOT using complete cases only and restricted to the anticoagulation domain \cite{mcquilten2023anticoagulation}. For simplicity we focused on comparing only the low vs intermediate dose with LMWH. Supplementary Material 4 summarises the distribution of falling in each category by treatment arm for each ordinal endpoint, which appear to mostly be skewed. Notably, both 29-point category outcomes have categories with either no or very few patients. 

When the five analysis methods were applied to the mMRC breathlessness scale, all methods estimated the posterior median cut-point log-OR of the ordinal scale to be smaller than zero for all cut-points. This implies that the low-dose provided better odds of a more favourable outcome, although there is considerable uncertainty given the credible intervals overlap with zero. The exception was the log-OR that corresponds to the last cumulative logit, where the unconstrained PPO, separate logistic regression and a constrained PPO model that allowed the last cumulative logit to diverge from PO all estimated an $\mathrm{OR} > 1$ (Figure 5). However, there is considerable uncertainty in these estimates as indicated by the $95\%$ credible intervals which, due to their overlap across each cumulative logit, also suggest that the PO assumption may be reasonable. The PO model produced the most efficient estimates compared with the all other models, particularly for the last cut-point, likely due to the smaller event rate in the upper categories.

\begin{figure}[h!]
\centering
\caption{Posterior median and 95\% credible intervals of cut-point log-ORs for the modified Medical Research Council five-point breathlessness scale in the Australasian COVID-19 Trial obtained under the five analysis methods}
\small\textsuperscript{OR = odds ratio, PO = proportional odds, PPO = partial proportional odds, LR = logistic regression, CPPO = constrained partial proportional odds}

\advance\leftskip-3cm
\advance\rightskip-3cm
\includegraphics[width=1.1\textwidth]{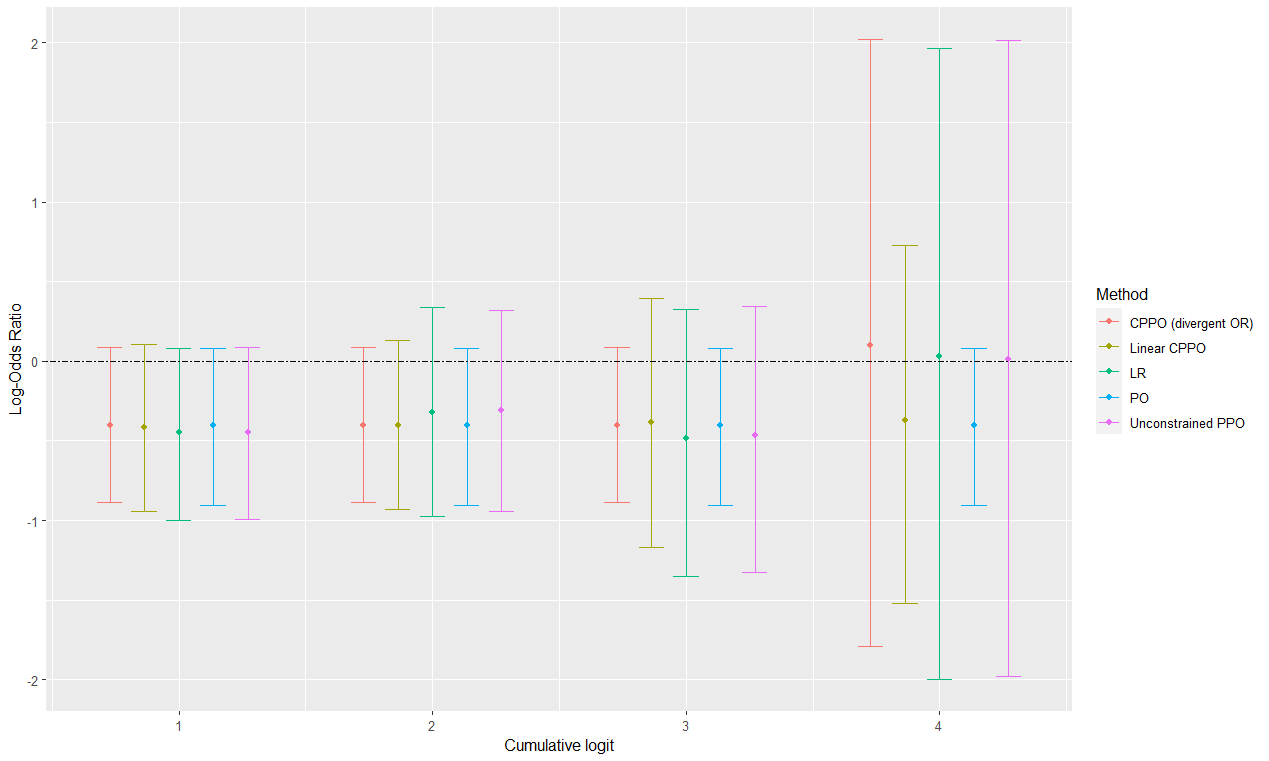}
\end{figure}

When the five analysis methods were applied to the remaining three endpoints, it appeared that the PO assumption was reasonable based on the point estimates and overlapping $95\%$ credible intervals estimated using separate logistic regression models for each cut-point (Figure 6 and Supplementary Material 4). Interestingly, the unconstrained PPO model estimated obscure cut-point log-ORs in cases where there were very few or no observed patients in each category for all endpoints (particularly when only observed in one treatment arm), which is likely due to the difficulty estimating the increment associated with the $k{th}$ cumulative logit, $\gamma_{k}$, due to the paucity of data. We note that there were a number of divergent transitions for this analysis when the unconstrained PPO model was used, indicating that the sampler could not explore the parameter space entirely despite increasing the maximum tree-depth and target average acceptance probability, which could also provide an explanation for these results.

\begin{figure}[h!]
\centering
\caption{Posterior median and 95\% credible intervals of cut-point log-ORs for the 8-point World Health Organisation clinical progression scale in the Australasian COVID-19 Trial obtained under the five analysis methods}
\small\textsuperscript{OR = odds ratio, PO = proportional odds, PPO = partial proportional odds, LR = logistic regression, CPPO = constrained partial proportional odds}

\advance\leftskip-3cm
\advance\rightskip-3cm
\includegraphics[width=1\textwidth]{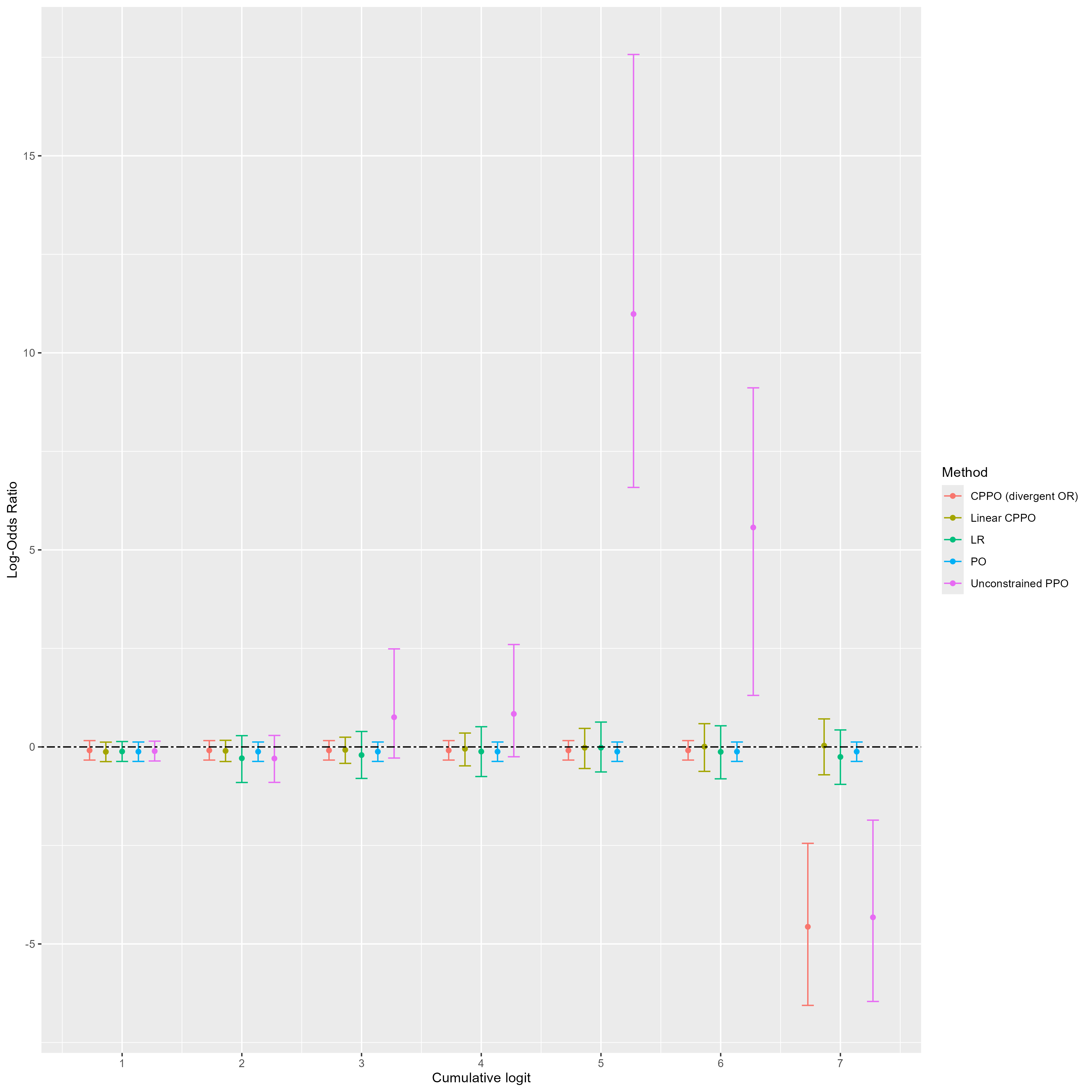}
\end{figure}

\section*{Discussion}
In this paper we evaluated the performance of separate logistic regression models, the PO model, and the unconstrained and constrained PPO models when PO does and does not hold. Simulation results showed that the PO model outperformed the other models in terms of bias and precision when the data generating mechanism used PO (as expected), but can result in substantial bias and under-coverage in the presence of non-PO, particularly at the extremes of the ordinal scale. In contrast, the cut-point ORs estimated using the unconstrained PPO and separate logistic regression models had negligible bias and good coverage in the presence of both PO and non-PO, with little difference between the performance of the two methods. Compared to all other models, the constrained PPO was more efficient and less biased, especially at the extreme ends of the ordinal scale, if the cut-point ORs were generated as a function of the ordinal scale cut-points. However, this approach relies on specifying the correct functional relationship which may be difficult to correctly specify in practice without existing data from pilot studies or related trials. 

When specifying the target estimand(s) in a trial, researchers should think carefully whether they are interested in a global (proportional) OR or separate ORs for each, or a single, cut-point. Misalignment of the target estimand compared to the analysis can lead to misinterpretation or loss of clinical relevance. Estimand specification should therefore prioritise meaningful cut-points on the scale, align with trial objectives, and ensure transparency in how effects are analysed and reported. By addressing potential non-proportionality in treatment effects during trial design, researchers can enhance the interpretability and clinical utility of their findings.

If using a PO model to estimate a global OR, it is important to check the proportionality of the ORs following data collection. A review of trials that used ordinal outcomes found that the PO model is commonly used in practice to estimate a treatment effect, but details on whether or how the PO assumption was assessed was lacking in the majority of these studies \cite{selman2024statistical}. If the PO assumption was found to be violated, studies either resorted to alternative statistical methods, such as using a Wilcoxon test without reporting a target parameter, dichotomised the ordinal scale, or dropped the variable that violated PO if it was not the treatment indicator. Only two studies in the review used an unconstrained PPO model both of which used this model as a sensitivity analysis, indicating that these methods are not often used in practice. 

Assessing the likely validity of the PO assumption could be considered prior to the commencement of trial recruitment using pilot or external datasets, and in the analysis stage,by plotting the ORs for each cut-point of the ordinal scale with the corresponding credibility (or confidence) interval estimated using separate logistic regression models (as in Figure 5). If the point estimates of the ORs all favour the same treatment and the CIs overlap, then making the PO assumption may be reasonable. If there is, however, variation in the point estimates across each cut-point of the scale such that the treatment appears both beneficial and harmful for different categories, or there are non-overlapping CIs, then there should be careful consideration as to an appropriate statistical approach to analyse the ordinal outcome. This could involve a discussion with collaborators to determine an appropriate and clinically meaningful estimand and whether the ordinal outcome could or should be dichotomised. The target estimand(s) and statistical approaches should be pre-specified in a statistical analysis plan before data analysis has begun, and if using the PO model, could indicate the alternative approach that will be used if the PO assumption appears to be violated. The PO assumption is analogous that used in the analysis of a time-to-event outcome that typically involves estimating a hazard ratio using a proportional hazards model that assumes that the hazard is constant across time. Similarly in this context, alternative analysis methods can be pre-specified if this assumption appears to be violated. 

The simulation study suggested that the unconstrained PPO model may offer a useful workaround when the PO assumption is violated, however the case study highlights that the unconstrained PPO model may fail to accurately estimate the cut-point ORs when there are very few or no observations that fall in the categories of the ordinal outcome which may be common in RCTs with small sample sizes or ordinal outcomes with a high number of categories. One possible reason for this is that we set the standard deviation associated with the non-PO component of the unconstrained PPO model to be large to ensure it was non-informative (and to be consistent with the same standard deviation used in the simulations), although this puts high prior probability on extreme treatment effects which may have been influential. It may be that reducing the standard deviation of the prior may improve the reliability of the results from the unconstrained PPO model. However, we note that reducing the standard deviation of the prior for the non-PO component closer to zero increasingly assumes PO. We explored this scenario in a couple of scenarios in the simulation study post-hoc to determine the effect of using a prior with smaller standard deviation. We found that using a prior standard deviation of $\sigma = 1$ or $10$  resulted in bias in the estimation of the cut-point proportional ORs when an unconstrained PPO model was used in scenarios where there were non-PO. The alternative approach of fitting separate logistic regression models performed well and does not suffer from these issues that we describe here, which should be used to estimate each cut-point OR in such cases.

In the current manuscript we have shown that the choice of analysis model is less critical when there is a small compared to a large number of categories in the ordinal outcome. For researchers and clinicians, this finding implies that in ordinal outcomes with a small number of categories, such as the four-point Glasgow Coma Scale, the PO model may be appropriate even if the PO assumption is violated. If the are a large number of categories and the PO assumption is violated, fitting separate logistic regression models to each cut-point or fitting an unconstrained PPO model and useful alternatives and will provide similar results (provided that there is not data sparsity). Given the widespread familiarity of the standard logistic regression model, this would seem a more natural choice. The use of a fully unconstrained or partially constrained specification (such as allowing the top category to vary freely from the PO assumption) may be beneficial when it is important to adjust for other covariates such as randomisation stratification factors if the stratification factors are assumed to satisfy PO. However, if the interest is in the dichotomisation at the extreme cut-point(s) of the ordinal scale and data exists that could guide the specification of a relationship between the cumulative log-ORs and cumulative logit, the constrained PPO model may improve efficiency (assuming it is correctly specified). 

A strength of this work is the investigation of a range of scenarios reflecting realistic scenarios (162 in total). This study has demonstrated that unconstrained and constrained PPO models offer a useful alternative modeling approach for ordinal outcomes which is scarcely used in practice. However, we have also shown that these models do not out-perform using separate logistic regression models which are a much more familiar approach. This study does, however, have its limitations. First, additional non-proportionality scenarios and adjustment for covariates, such as stratification factors, were not considered. Other scenarios exist in reality, such as more categories in the outcome, there being a U-shape or highly skewed relationship between the cumulative logits and cut-point log-OR, and there being close to zero probabilities in some of the outcome categories, although the case study provides some insight regarding some of these conditions. We also restricted the analysis of interest to the Bayesian setting, although we expect similar findings using frequentist methods given that we have used vague priors, although this could be an area for future work. The standard deviation we specified for the prior on the treatment effect was also quite large ($\sigma = 100$). While this prior seems reasonable on the log-odds scale, applying the inverse-logit transformation shifts the distribution's mass toward the extremes near zero and one, resulting in a `bathtub'-shaped distribution. However, we did explore some scenarios post-hoc where we reduced the value of the standard deviation and did not find a difference in most results, with the exception of the unconstrained PPO model in which some bias was introduced. This is because a smaller standard deviation for the non-PO parameter essentially assumes PO even in the presence of non-PO. Finally, the performance of the methods were not evaluated in settings with smaller sample sizes. As described above, the decision to use larger sample sizes was to ensure the posterior distribution was data driven rather than highly influenced by the priors. It is currently unclear what an appropriate non-informative prior specification is for the different PO and PPO models making it hard to make comparisons between the inference from the  models in small sample sizes. Evaluating the impact of different prior specifications on the inference from the PO model is beyond the scope of this paper and is being investigated in a separate project by the co-authors.  

In summary, although the PO model is a useful approach if either the PO assumption is somewhat reasonable  or if there are few categories in the ordinal outcome, departures from the PO assumption can result in invalid conclusions about the treatment effect, especially when the treatment has varying impact across different levels of the ordinal scale. Separate logistic or PPO models may provide flexible alternatives to the PO model where the PO assumption does not hold, however their use requires redefinition of the target estimands, should make sense in terms of the objectives of the analysis, and should ideally be pre-specified.

\newpage 

%%%%%%%%%%%%%%%%%%%%%%%%%%%%%%%%%%%%%%%%%%%%%%
%%                                          %%
%% Backmatter begins here                   %%
%%                                          %%
%%%%%%%%%%%%%%%%%%%%%%%%%%%%%%%%%%%%%%%%%%%%%%

\begin{backmatter}
\section*{Declarations}
\subsection*{Acknowledgements}%% if any
The authors thanks the Ackman Trust (UTR6.170) that awarded The Henry and Rachael Ackman travelling scholarship as well as the Australian Trials Methodology (AusTriM) research network in order to present this work at an overseas conference. The authors also thank the Australian COVID-19 (ASCOT) trial team for providing access and use of the ASCOT trial data to illustrate the methods discussed in this paper. 

\subsection*{Funding}%% if any
This work forms part of Chris Selman’s PhD, which is supported by the Research Training Program Scholarship, administered by the Australian Commonwealth Government and The University of Melbourne, Australia. Chris Selman's PhD was also supported by a Centre of Research Excellence grant from the National Health and Medical Research Council of Australia ID 1171422, to the AusTriM Research Network. Research at the Murdoch Children’s Research Institute is supported by the Victorian Government’s Operational Infrastructure Support Program. This work was supported by the Australian National Health and Medical Research Council (NHMRC) Centre for Research Excellence grants to the Victorian Centre for Biostatistics (ID1035261) and the Australian Trials Methodology Research Network (ID1171422), including through seed funding awarded to Robert Mahar. Katherine Lee is funded by an NHMRC investigator grant (ID2017498). The funding bodies played no role in the study conception, design, data collection, data analysis, data interpretation or writing of the report.

\subsection*{Consent for Publication}%% if any
Not applicable.

\subsection*{Availability of data and materials}
The code and datasets generated for this simulation study are available on Github \cite{SelmanGit2024}: \url{https://github.com/chrisselman/propodds}.

Data used for the illustrative example within this paper are available upon reasonable request from the ASCOT study team.
%% Need to add statement about ASCOT data 

\subsection*{Abbreviations}%% if any
ASCOT: Australasian COVID-19 Trial, ESS: Effective Sample Size, HMC: Hamiltonian Monte Carlo, LMWH: Low-Molecular-Weight Heparin, RCT: randomised controlled trial, OR: odds ratio, PO: proportional odds, PPO: partial proportional odds, MSE: mean square error, MCSE: Monte Carlo standard error, mMRC: Modified Medical Research Council, WHO: World Health Organisation. 

\subsection*{Ethics approval and consent to participate}%% if any
The ASCOT trial had ethics approval (detailed separately in the reference to the trial manuscript). This study had approval to use the data for illustration as a case study.

\subsection*{Competing interests}
The authors declare that they have no competing interests.

\subsection*{Authors' contributions}
CJS, KJL, RKM conceived the study. CJS conducted the simulation study and data analysis and wrote the first draft of the manuscript, with regular input from KJL and RKM. All authors contributed to the design of the study, critical revision of the manuscript, and take responsibility for its content.

%%  After submission of the .TEX file,                     %%
%%  you will be prompted to submit your .BBL file.         %%

% if your bibliography is in bibtex format, use those commands:
\bibliographystyle{bmc-mathphys} % Style BST file (bmc-mathphys, vancouver, spbasic).
\bibliography{bmc_article}      % Bibliography file (usually '*.bib' )
% for author-year bibliography (bmc-mathphys or spbasic)
% a) write to bib file (bmc-mathphys only)
% @settings{label, options="nameyear"}

% or include bibliography directly:
% \begin{thebibliography}
% \bibitem{b1}
% \end{thebibliography}

\end{backmatter}

\end{document}